\documentclass[aps,prb,twocolumn,10pt,longbibliography,floatfix,superscriptaddress]{revtex4-2}

\usepackage{physics}
\usepackage[english]{babel}
\usepackage[utf8]{inputenc}
\usepackage{amssymb}
\usepackage{amsmath}
\usepackage{dsfont}
\usepackage{multirow}
\usepackage{tabularx}
\usepackage{units}

\usepackage[pdftex]{graphicx}
\usepackage{xspace}
\usepackage{dsfont}
\usepackage{bm}
\usepackage[pdfencoding=auto, psdextra]{hyperref}
\hypersetup{
    colorlinks,%
    citecolor=blue,%
    filecolor=blue,%
    linkcolor=blue,%
    urlcolor=blue
}

\usepackage[usenames, dvipsnames]{xcolor}

\begin{document}

\title{sp\texorpdfstring{$^{2}$}{TEXT}/sp\texorpdfstring{$^{3}$}{TEXT} bonding controlling mechanism at the \texorpdfstring{$\alpha$}{TEXT}-Al\texorpdfstring{$_{2}$}{TEXT}O\texorpdfstring{$_{3}|$}{TEXT}graphene interface}

\author{Renan~P.~Maciel}
\affiliation{Department of Physics and Astronomy, Division of Materials Theory, Uppsala University, Box 516, SE-75120 Uppsala, Sweden}

\author{Chin Shen Ong}
\affiliation{Department of Physics and Astronomy, Division of Materials Theory, Uppsala University, Box 516, SE-75120 Uppsala, Sweden}

\author{Daria Belotcerkovtceva}
\affiliation{Department of Materials Science and Engineering, Uppsala University, Box 35, SE-751 03 Uppsala, Sweden}

\author{Yaroslav O. Kvashnin}
\affiliation{Department of Physics and Astronomy, Division of Materials Theory, Uppsala University, Box 516, SE-75120 Uppsala, Sweden}

\author{Danny Thonig}
\affiliation{School of Science and Technology, \"Orebro University, Fakultetsgatan 1, SE-70182 \"Orebro, Sweden}
\affiliation{Department of Physics and Astronomy, Division of Materials Theory, Uppsala University, Box 516, SE-75120 Uppsala, Sweden}

\author{M. Venkata Kamalakar}
\affiliation{Department of Materials Science and Engineering, Uppsala University, Box 35, SE-751 03 Uppsala, Sweden}

\author{Olle Eriksson}
\affiliation{Department of Physics and Astronomy, Division of Materials Theory, Uppsala University, Box 516, SE-75120 Uppsala, Sweden}
\affiliation{School of Science and Technology, \"Orebro University, Fakultetsgatan 1, SE-70182 \"Orebro, Sweden}

\date{\today}


\begin{abstract}
First-principles calculations reported here illuminate the effects of the interfacial properties of $\alpha$-Al$_{2}$O$_{3}$ and graphene, with emphasis on the structural and electronic properties. Various contact interfaces and different $\alpha$-Al$_{2}$O$_{3}$ surface terminations are considered with on and slightly-off stoichiometric aluminium oxide. We show that depending on whether aluminium or oxygen is in contact with graphene, an $sp^{3}$ structural deformation and spontaneous spin-polarization may occur next to the interface contact. Interestingly, some cases cause a $p$-type doping in the graphene band structure, depending on the initial $\alpha$-Al$_{2}$O$_{3}$ geometry placed on graphene. The importance of leaving the surface dangling bonds of alumina saturated or not is also highlighted, and we show that it might be a control mechanism for opening a gap in graphene by the influence of the $sp^{3}$ bond between oxygen and carbon atoms at the interface. We discuss the potential of utilizing this sensitivity for practical applications.
\end{abstract}
\maketitle
\section{Introduction}
\begin{figure*}[t!]
  \centering
  \includegraphics[width=1.0\textwidth]{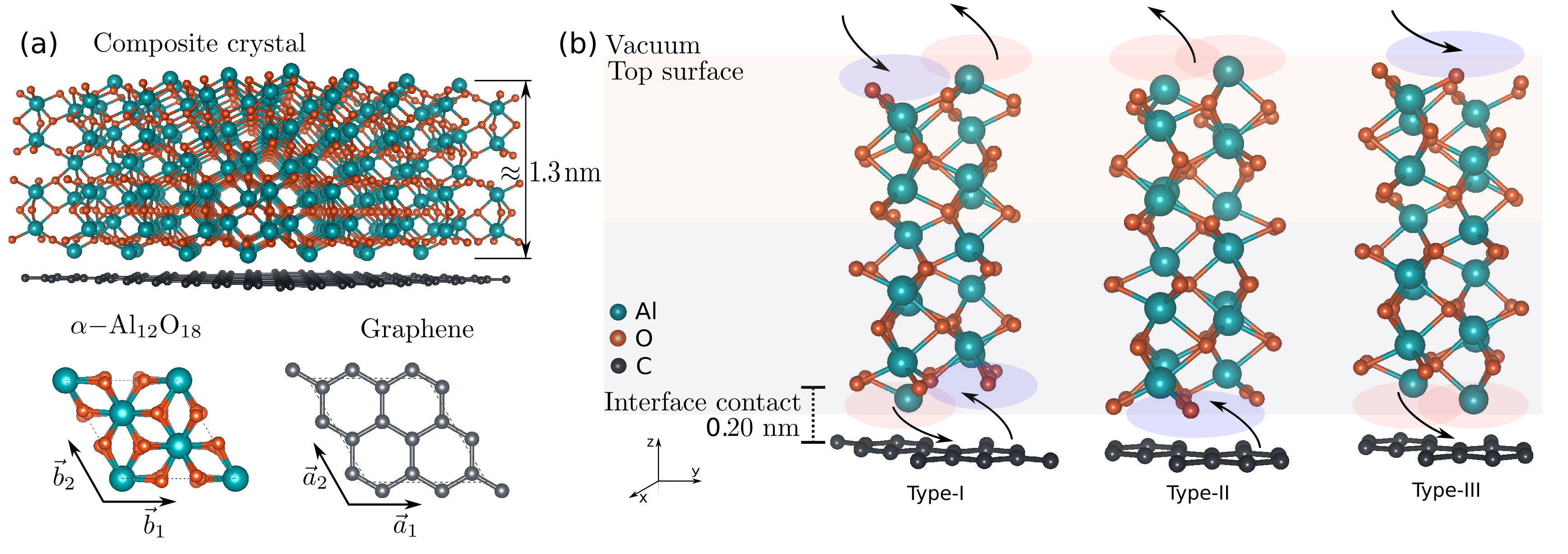}
  \caption{ A schematic illustration of the unrelaxed slabs. (a) Al$_{12}$O$_{18}$ thin film on top of the graphene monolayer. The thickness of the film and an illustration of the transformation from rhombohedral to a hexagonal cell is also shown. (b)  Hexagonal Al$_{12}$O$_{18}$ slab with three different surface/contact interface contact, namely, type-I, type-II, and type-III. Red and blue regions highlight the unsaturated Al and O bonds, respectively. The arrows represent when the atoms are electron donors or receptors.}
  \label{fig:model}
\end{figure*}

Graphene is an attractive semi-metallic two-dimensional material that has attracted attention both from experiment and theory due to its remarkable electronic and mechanical properties~\cite{katsnelson2007graphene}. This involves high carrier mobility, resilience against strain, vanishing density of states (DOS) at the Fermi level, linear electronic band dispersion (Dirac cone), and high chemical stability. Many theorists have investigated these proprieties using first-principle electronic structure calculations, tight-binding theory, or analytical expressions that are possible due to the relatively simple linear Hamiltonian with a momentum dependence that accomplishes most of the graphene fundamental physics~\cite{katsnelson2007graphene,geim2010rise, wallace1947band, mcclure1956diamagnetism, neto2009electronic}. This has provided a solid understanding of the most interesting features involving graphene and has allowed the design of new technologies, e.g., as a base for new and improved electronic devices.
Theoretical investigations of graphene show that the intrinsic carrier mobility might reach values up to the order of $10^{5}$~cm$^{2}$/(V$\cdot$s) \cite{morozov2008giant, bolotin2008ultrahigh}, and such values make graphene very promising for high-speed devices. However, a long-standing problem with graphene electronics has been the difficulty of introducing significant differences in conductivity as a function of gate voltage due to its gapless electronic structure and the Klein tunneling phenomena. 

Regardless of the remarkable advancement in demonstrating the potential of graphene for graphene electronics and spintronics, it remains a great challenge to optimize conditions that can preserve graphene's intrinsic electrical properties in device structures. For instance, carrier mobility in graphene varies dramatically whether it is suspended~\cite{xu2008approaching} or it is on atomically flat substrates~\cite{dean2010boron} or regular Si/SiO2 substrates~\cite{panda2020ultimate}. Nevertheless, several recent studies in molecular beam epitaxy (MBE), lithography, and atomic layer deposition (ALD) have shown substantial experimental improvement, allowing for more plausible matching between experiment and theoretical predictions. 
Of particular importance for the current investigation, one should mention that the key to most of these experimental techniques also relies on 
an understanding of how the interaction between graphene and the different substrates or electrode interfaces might interfere with the spin and charge transfer process that boosts the performance of graphene-based devices. These interactions play a central role in composite structures. Particularly, it is equally important to investigate whether these interactions might cause a gap opening in the graphene electronic structure. Therefore a detailed description of many different interactions has been explored in the literature. For example, Giovannetti et al.~\cite{giovannetti2008doping} studied the electronic structure of graphene when adsorbed on a variety of metallic surfaces (e.g., Al, Co, Ni, Cu). Their findings reveal that depending on the difference in work functions of the substrate and the adsorbed carbon lattice, graphene can be doped as either $n$-type or $p$-type. 

In the present work, we study the structural and electronic effects of the $\alpha$-Al$_{2}$O$_{3}$, as an ultra-thin layer ($\approx$ nm thickness ) in contact with the graphene lattice. This is because developing such an understanding is of practical significance as aluminium oxide is widely used in graphene electronics as gate dielectrics, effective barriers for graphene tunnel junctions~\cite{feng2016graphene}, and thin-film memristive devices~\cite{vu2017high} and resistive switching phenomena~\cite{huang2017graphene}. In addition, for both planar and vertical graphene spintronic devices~\cite{dayen2020two, godel2014voltage}, ultra-thin layers of the aluminium oxide are employed as tunnel barriers to achieving optimum interface for efficient electrical spin injection~\cite{dlubak2012highly, han2010g}. For instance, Amamou \textit{et al.} have pointed out the importance of tunneling properties and spin relaxation process in graphene spin valve devices~\cite{amamou2016contact}. However, to the best of our knowledge, 
either several theoretical work focus on the isolated properties of $\alpha$-Al$_{2}$O$_{3}$~\cite{french1970composition, causa1989ab, godin1994atomic, puchin1997atomic, ahn1997composition} or experimental analysis investigates the interface conditions of the $\alpha$-Al$_{2}$O$_{3}|$graphene system~\cite{amamou2016contact, feng2016graphene,  li2014uniform,iqbal2017enhanced,tateno2020investigation}. Therefore, it is important to have a bridge between these investigations that allow us to have a deeper insight into the electronic and structural properties of the $\alpha-$Al$_{12}$O$_{18}|$graphene interface.

The $\alpha$-Al$_{2}$O$_{3}$ has often been suggested as a suitable substrate because of its large electronic bandgap. Experiments show that the gap can vary from $\sim 5.0$ to $\sim \unit[8.8]{eV}$ depending on the crystal phase and how the oxide is synthesized~\cite{filatova2015interpretation, bharthasaradhi2016structural}. Regardless of its growth condition and various crystal phases, the primary purpose of this research is to demonstrate that different contact conditions between $\alpha$-Al$_{2}$O$_{3}$ and graphene can produce significantly varied properties. Furthermore, we examine the implications of saturating the dangling bonds at the top surface layer of aluminium oxide, also known as alumina, of the many terminations that the thin film can present once it can drastically influence the graphene band structure, primarily depending on whether the dangling bonds are saturated or not. Our investigation is mainly based on mimicking the experimental configuration reported recently~\cite{belotcerkovtceva2022insights} and helps to clarify the electronic phenomena involving $\alpha$-Al$_{2}$O$_{3}$ with device structures of experimental significance, in particular to interfaces in graphene electronic tunnel junctions~\cite{feng2016graphene}, memristors~\cite{vu2017high}, and planar and vertical graphene spintronics devices~\cite{dayen2020two, godel2014voltage}.

The paper organization is as follows. Sec.~\ref{sec:model} summarizes the computational details used in this work. In addition, we discuss how we build the model and the system geometry. Sec.~\ref{sec:results} shows the main results related to the various $\alpha$-Al$_{2}$O$_{3}|$graphene interfaces. We also study the effects of applying strain when the slab is created and discuss the effects of the on and slightly-off stoichiometric aluminium oxide on graphene band structure when dangling bonds are saturated or not.

\section{The model}\label{sec:model}

\subsection{Computational details}
All calculations were done using the Vienna ab-initio simulation package (VASP)\cite{kresse1993ab, kresse1996efficiency, kresse1996efficient} using the Projector augmented wave method with Perdew–Burke–Ernzerhof (PAW-PBE) exchange-correlation functionals~\cite{ kresse1999ultrasoft}. Due to this functional underestimating atom bindings, we used the van der Waals correction by applying the Becke-Jonson damping method~\cite{grimme2011effect,schr2015reformulation} when simulating the combined oxide and graphene structure. In addition, the plane-wave basis set was chosen with an energy cutoff of $\unit[520]{eV}$ for the kinetic energy, a mesh of $\mathbf{k}$-points of $6\times6\times1$ subdivisions was employed, and a dipole correction is also considered. The spin and non-spin polarization results are obtained by performing a self-consistent calculation in the optimized structure. 

\subsection{The slab construction}

In this section, we describe the models used in this paper. We start by matching the $\alpha$-Al$_{2}$O$_{3}$ $1\times1$ hexagonal crystal structure with lattice vectors $|\textbf{b}_1|=|\textbf{b}_2|\approx\unit[0.476]{nm}$ with graphene 2$\times$2 supercell with lattice vectors $|\textbf{a}_1|=|\textbf{a}_2|\approx\unit[0.492]{nm}$. The lattice mismatch between both structures is $\epsilon=3.25\%$. Due to this mismatch, a small strain must be applied to construct the slab. We have chosen to strain the oxide instead of compressing graphene to ensure that all related electronic properties of graphene are mainly preserved, and the only influence is due to the interaction with $\alpha$-alumina in a slightly compressed state. We will further show in Sec.~\ref{sec:resultsA}. that the main conclusions of this work are not affected by straining either graphene of alumina in the calculations.

For the slab, a semi-infinite system with $\approx\unit[0.13]{nm}$ thickness along the $z$-axis was considered, correspondent to the stoichiometry Al$_{12}$O$_{18}$. It is worth noting that the slab may show different surface terminations (i.e., oxygen or aluminium as the top layer of Fig.\ref{fig:model}(a)) as well as different terminations in the interface contact with the graphene layer. For this reason, we label each different geometry as type-I, $-II$, and $-III$. The type-I indicates the oxide terminations present $Al$ in both endings. The type-II is the configuration where $Al$ atoms are next to the graphene layer while $O$ is at the top surface, and type-III is the extreme opposite of the type-II. A general view of the slabs and the Al$_{12}$O$_{18}|$C$_{8}$ interfaces is shown in Fig.~\ref{fig:model}(b). Particular attention was paid to the initial distance between graphene and the oxide layer since this might enhance or diminish the chemical interaction at the crystal interface, and one may end up in different (meta) stable solutions depending on this initial choice. For this reason, we set up an initial value of $\approx \unit[0.2]{nm}$ for all models studied in this work before structural optimization. This chosen distance is based on the convergence of the energy minima after several analyses considering different initial distances within a range of $0.1$ to $\unit[0.35]{nm}$. From such calculations, three distinct geometries might be found by shifting the perfect stoichiometric oxide $\alpha$-Al$_{12}$O$_{18}$ hexagonal unit cell plane in the \{0001\} direction. Each different surface termination is related to a dangling bond from either Al or O (or both simultaneously) that can emerge at the surface of the aluminium oxide layer or the interface with graphene.

Concerning the coordination number of the aluminium oxide, one might notice that in the bulk structure, Al atoms are connected with six O nearest-neighbor atoms while four Al atoms surround O. This means each trivalent Al atom contributes with $1/2$~$e$(where $e$ is the electron charge) electron bond, while hexavalent O contributes with $3/2$~$e$. Conversely, because of the broken symmetry along the $z$-direction, the slab structure will hold a reduced coordination number at the top/bottom surfaces for both Al and O atoms. For instance, the aluminium oxide in the type-I system in Fig.\ref{fig:model}(b) illustrates when Al and O atoms have their bonds reduced to three at the top and interface contact, and consequently, $1/2$~$e$ is donated (absorbed) by aluminium (oxygen) atoms. On the other hand, type-III and type-II structures, the Al (O) atoms and donate (absorbs) $3$~$e$. It is worth recalling that a similar number of dangling bonds (independently of the alumina geometry) leads to similar surface interactions. Additionally, to have further insight into the electronic structure of graphene by the influence of AlO$_{x}$, grown under different experimental conditions, we also consider a slightly off-stoichiometry by removing either one Al or O atom from Al$_{12}$O$_{18}$ slab (i.e., we consider Al$_{11}$O$_{18}$ and Al$_{12}$O$_{17}$, respectively) at the interface between the oxide and graphene layer. 

\section{Results}\label{sec:results}

\begin{figure}[b!]
  \centering
  \includegraphics[width=1.0\columnwidth]{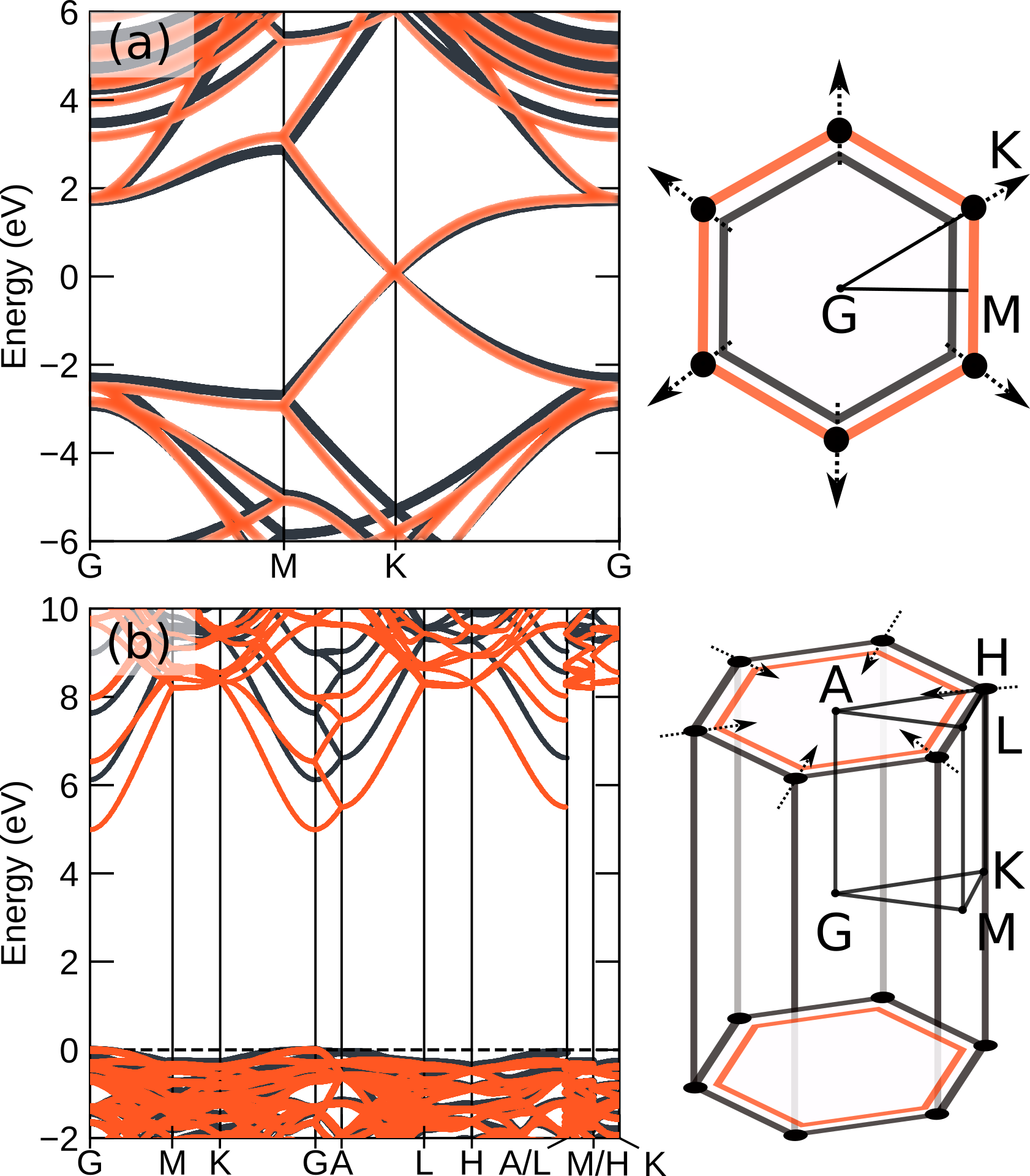}
  \caption{Superposition of the unstrained (black) and strained (orange) band structure for (a) graphene monolayer with compression of $\approx 3\%$, and for (b) strained hexagonal bulk $\alpha-$Al$_{12}$O$_{18}$. At the right-hand side of each band structure, we schematically show the Brillouin Zone before and after applying compression/strain.}
  \label{fig:bs_strained_no_strained}
\end{figure}

\subsection{Strained vs unstrained system}\label{sec:resultsA}

We first studied the effects of strain on alumina  (under tensile strain) and the 2$\times$2 graphene supercell  (under compressive strain) on their respective electronic band structures before moving on to the composite Al$_{12}$O$_{18}|$graphene slab. To perform this analysis, we exchanged the lattice parameters between both crystals so that the strained alumina had a lattice constant of $\approx\unit[0.492]{nm}$, while for graphene, it was $\approx\unit[0.476]{nm}$. To take into account the mentioned strain, we fixed the new unit cell lattice parameter and relaxed all atomic positions of each system such that individual lattice forces are less than $\unit[10^{-4}]{eV/nm}$. After relaxation, we found new bond lengths for both structures. For graphene, the new in-plane C-C bond was $\unit[0.137]{nm}$, and in the case of alumina, the bond lengths between Al and O atoms increased from $1.86$ and $\unit[1.97]{nm}$ to $1.91$ and $\unit[2.01]{nm}$, respectively. 

The obtained band structure for each case is shown in Figs.~\ref{fig:bs_strained_no_strained}(a)-(b). The figure shows that the applied compression of the graphene supercell is not enough to change its band structure significantly. The reduction of approximately $\approx 3\%$ of carbon-carbon bond length is insufficient to distort the linear dispersion at the Fermi level or to produce a gap by violating either time-reversal of space inversion symmetry, which protects the presence of Dirac cone, and therefore, the system remains gapless~\cite{sahu2017band}. This behavior is expected, as discussed by several authors in Refs.\cite{neto2009electronic,sahu2017band}. However, the increased overlap between in-plane C orbitals induces a slightly modified overlap between in-plane $sp^{2}$ orbitals. Consequently the $p_{x}$, $p_{y}$ orbitals from carbon atoms in both conductance  and valence bands are shifted as shown in  Fig.\ref{fig:bs_strained_no_strained}(a). The figure shows a larger splitting between bonding and anti-bonding states when the orbital overlap increases.

For alumina, the Al-O bond increases, which induces a bandgap reduction from $\unit[6.12]{eV}$ to $\unit[5.00]{eV}$. This is caused by a rigid shift of the valence band and conduction band states. If one chooses to align the top of the valence bands (as is done in Fig.\ref{fig:bs_strained_no_strained}(b)), this shift can be described as pushing the conduction band down while the valence band states (composed of oxygen $p-$orbital states) are mostly unchanged. From this analysis, we conclude that a study of the electronic structure of the Al$_{12}$O$_{18}|$graphene system with a common in-plane lattice constant can be done either by adapting the C lattice constant to alumina or vice versa. The main conclusion will not be influenced by this choice, and here we have chosen to construct the Al$_{12}$O$_{18}|$graphene interface by adapting the oxide layer to the graphene unit cell. 

\subsection{Al\texorpdfstring{$_{12}$}{TEXT} O\texorpdfstring{$_{18}|$}{TEXT}graphene interface}

\begin{figure}[t]
  \centering
  \includegraphics[width=1.0\columnwidth]{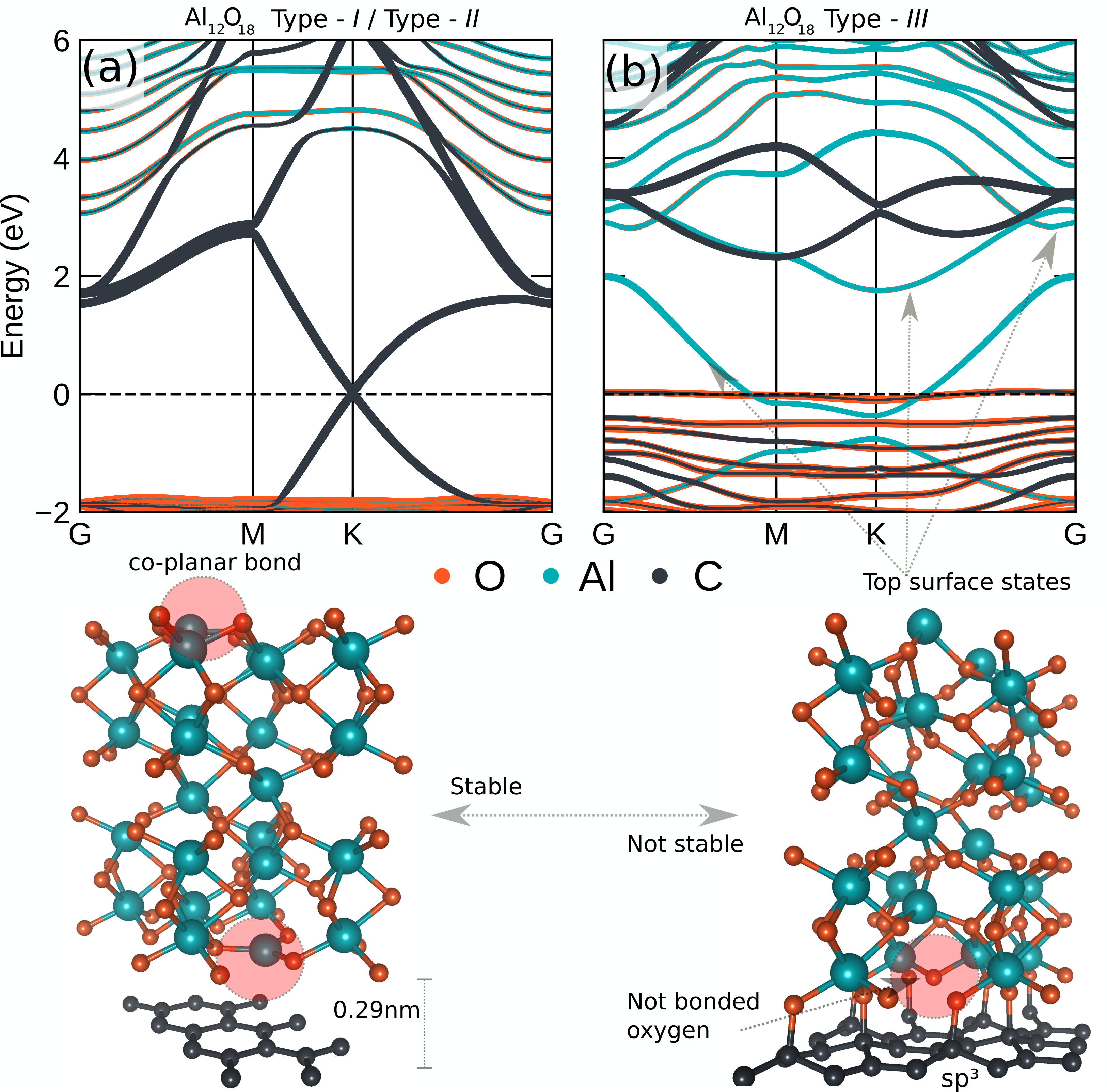}
  \caption{Atom projected band structure of the Al$_{12}$O$_{18}|$graphene structure where the color code corresponds to the atom projection. (a) Shows the Dirac cone centered at the Fermi Level of its representative geometry that is right below. This picture illustrates the geometric configuration of the type-I and type-II after all atoms are relaxed, and the circles highlight where the co-planar bond is formed. (b) Band structure of the type-III system. The grey arrows highlight the band states that crossing the Dirac cone and the Fermi level is due to the dangling bonds of the top surface layer of the oxide slab. This illustration also shows the emergent $sp^{3}$ bonds at the graphene-alumina interface and the oxygen atoms not bonded to graphene.}
  \label{fig:C-Al12O18-NS-bs}
\end{figure}

To investigate the influence of the aluminium oxide on the graphene electronic structure, we first consider a graphene substrate with type-I, -II, and -III $\alpha$-Al$_{12}$O$_{18}$ geometries on top (see Fig.~\ref{fig:model}(b)). Secondly, we relax the atomic positions of each case, and finally, we discuss the main results regarding the stability of the different structures. In the discussion below, we will refer to Al and O atoms at the graphene (bottom) interface and the top surface. This means atoms in Fig.~\ref{fig:model}(b) that are either in contact with the graphene layer or are at the other end of the structure, forming the outermost surface layer. 

In type-I geometry, we found that the specific geometry of the aluminium oxide on top of graphene presents the same number of dangling bonds at the top surface and contact interface near graphene. After relaxing this structure, the Al atom interacts primarily with the oxygen in its nearest surrounding and forms a coplanar bond avoiding interacting strongly with the graphene layer underneath (see (Fig.~\ref{fig:C-Al12O18-NS-bs}(a)). This can be seen as an auto compensation of charge transfer, where similar to the bulk environment, the charge is donated from Al to O, implying charge neutrality of the oxide slab, as indicated by the red and blue ovals in Fig.~\ref{fig:model}(b) (type-I structure). As a result, there is only a weak interaction between the oxide and graphene, which is confirmed by the increase of $\approx \unit[0.09]{nm}$ between these two slabs from its initial configuration. Nonetheless, no substantial change in the graphene electronic band structure is observed, such as doping or shifting of the Dirac cone (see Fig.~\ref{fig:C-Al12O18-NS-bs}(a)). In terms of the obtained total energy, we found that this system is the most stable structure of our analysis (see Table~\ref{table:al12o18} for a summary of energies of the different systems studied here).

Surprisingly, our result shows that the oxide of the type-II geometry also avoids interacting with graphene, even though its initial configuration has a different number of dangling bonds at both endings that do not compensate for each other from a charge transfer perspective (indicated by the red and blue regions on either side of the alumina in Fig.~\ref{fig:model}(b)). After relaxing this structure, we found the same total energy as in the previous case, i.e., the type-I geometry. These two systems were indeed found to relax to the same final geometry and show identical electronic structure (see Fig.~\ref{fig:C-Al12O18-NS-bs}(a)) and total energy (Table~\ref{table:al12o18}). This suggests that the atomic arrangement, charge redistribution, of the thin-film after the relaxation tend to neutralize the charge transfer process between the slabs once Al atoms are near graphene.  

The type-III system represents the less stable geometry. We obtained total energy of $\approx$~9.33 ~eV (per slab containing 38 atoms) higher than the type-I and $-II$ structures (see Table~\ref{table:al12o18}). Here, two out of three oxygen atom at the interface interacts with graphene and forms a C-O covalent bond, as is clear from the bond angles and bond distances are shown in the bottom of Fig.~\ref{fig:C-Al12O18-NS-bs}(b). (Note that each unit cell of a-alumina contains 6 formula units of Al2O3). Unlike the type-I and $-II$ systems, this structure does not show any coplanar bond after the relaxation. We observe, however, that at the top surface, the Al atoms remain undistorted concerning their initial configuration, while only the O atoms at the interface induce an out-of-plane deformation (forming $sp^3$ hybrid states) towards the graphene layer. This causes a strong hybridization between O and C atoms reflected in the graphene electronic band structure, as shown in top figure in Fig.\ref{fig:C-Al12O18-NS-bs}(b). As may be seen, the strong hybridization between O and C states distorts the entire electronic structure, hence removing the Dirac cone.

\begin{table}[!ht]
\caption{Summary of the energy and magnetic properties of the Al$_{12}$O$_{18}|$graphene slab. }
\begin{tabular}{l|c|c|c}
Al$_{12}$O$_{18}|$graphene & \multicolumn{1}{l|}{Type-I} & \multicolumn{1}{l|}{Type-II} & \multicolumn{1}{l}{Type-III} \\ \hline
$\Delta E_{t} $\footnote{Energy difference $\Delta E_{t}$ per  slab unit cell (s.u.c) relative to the most stable structure. } (eV)      & 0 & 0 & 9.33                        \\
$\mu_{B}$\footnote{Total magnetization $\mu_{B}$ per slab unit cell (s.u.c.) } & 0.00 & 0.00 & 1.17                       \\
$\mu_{B}$ of unbonded O atom                                                          & 0.00 & 0.00 & 0.32                          
\end{tabular}\label{table:al12o18}
\end{table}

\subsection{Slightly off-stoichiometric oxide}

\begin{figure*}[ht]
  \centering
\includegraphics[width=1.0\textwidth]{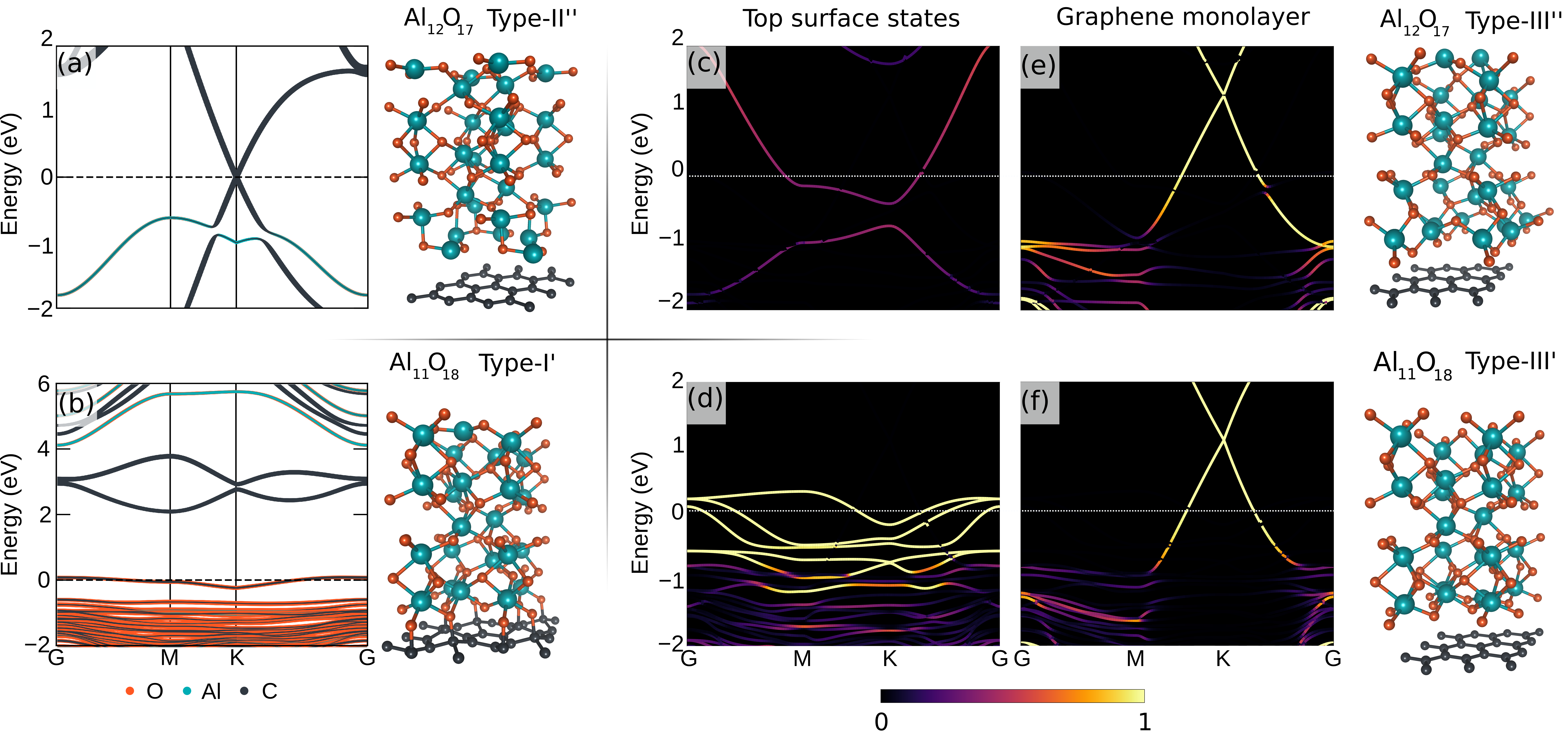}
\caption{Band structures of the slightly off-stoichiometric Al$_{12}$O$_{17}$ (upper panels) and Al$_{11}$O$_{18}$ (lower panels) system. On the left, we show the most sable representative structure of the sub-stoichiometric system next to their representative band structure. On the right is the less stable one. (a) The band structure of the Al$_{12}$O$_{17}$ type-II'' geometry shows the hybridization between carbon and aluminium $p$-orbital state. Here we also show the atomic structural position of the alumina composite after full relaxation. (b) Band structure of the Al$_{11}$O$_{18}$ type-I' shows a completely distorted Dirac cone, and the $sp^{3}$ bond formation between oxygen and a carbon atom. (c) Top surface states from Al$_{12}$O$_{17}$ type-III'' and (d) Al$_{11}$O$_{18}$ type-II'. Band structure of the graphene monolayer showing $p$-type doping due to the (e) Al$_{12}$O$_{17}$ type-III'' and (f) Al$_{11}$O$_{18}$ type-II'. The representative color of the atom projected band is the same used to plot the geometric structure. The color bar indicates the weight contribution of an atom set to that particular band, where zero (indicated by the black color) means no contribution.}
  \label{fig:C-Al11O18-Al12O17-bs}
\end{figure*}

The structural and electronic properties of the $AlO_{x}|$graphene system were also investigated by removing one of the aluminium atoms close to the interface with graphene while keeping the top surface intact. This scenario considers an experimental scenario when the aluminium is not fully oxidized or if there is an elevated oxygen concentration next to the graphene layer. Therefore, this section will focus on discussing an oxide with the stoichiometry Al$_{11}$O$_{18}$. We also show data for a system with decreased oxygen concentration, i.e., the system Al$_{12}$O$_{17}$, and present data on how it interacts with graphene. 
In each case, we repeated the procedure of force minimization and obtained the total energy of a relaxed geometry. Our starting points are based on the (three) geometries discussed in the previous section, and we keep the notation to refer which geometry the off-stoichiometric system was derived from, with the slight modification of using a prime (') or double prime ('') to indicate the off-stoichiometric case of one Al or O less, i.e., type-I', -II', -III', and type-I'', -II'', -III'', respectively.

\subsubsection{The Al\texorpdfstring{$_{12}$}{TEXT} O\texorpdfstring{$_{17}|$}{TEXT}graphene interface.} The reduced oxygen concentration at the interface implies a reduction of the local coordination number, which generates more dangling bonds near the graphene layer. The oxygen removal causes a more dominant charge imbalance in the newly constructed geometries compared to the stoichiometric cases. (It is worth mentioning that such atom removal is done randomly since all oxygen at the interface contact has similar coordination numbers). After atomic relaxation, the lack of oxygen forces the thin film to reconstruct its geometry next to the graphene layer (see crystal structure in Fig.~\ref{fig:C-Al11O18-Al12O17-bs} (a)), and unlike Al$_{12}$O$_{18}|$graphene cases, all Al$_{12}$O$_{17}|$graphene structures avoids covalent bond formation between the oxygen of the slab and graphene. This suggests that any interaction between the oxide and graphene might occur via van der Waals forces once there is an oxygen deficiency. The final distance between graphene and the oxide for all Al$_{12}$O$_{17}|$graphene geometries increases to $\approx \unit[0.3]{nm}$. In terms of stability, we found that the most stable structure is the geometry type-II''.

From the electronic structure perspective, the most stable geometry (i.e., type-II'' structure) does not show any significant change in the graphene band structure. However, the absence of an oxygen atom force the $p$-orbits from aluminium to partially hybridize with the $\pi$-orbital of graphene. This hybridization mainly modifies the valence bands below the Fermi level and does not shift or open a gap of the graphene Dirac cone, as it is shown in Fig.~\ref{fig:C-Al11O18-Al12O17-bs} (a). (A very similar result is found to the type-I' geometry that has not been shown here). 

In contrast, for the less stable geometry (i.e., type-III''), the oxygen atom interacts indirectly with the graphene layer. This implies that there is a charge transfer between oxygen and carbon atoms without any sp$^{3}$ or covalent bond formation that would cause structural distortion of graphene. Therefore, the Dirac cone shifts $\approx \unit[1.2]{eV}$ above the Fermi level, indicating significant $p$-type doping, as shown in Fig.~\ref{fig:C-Al11O18-Al12O17-bs} (e). Note that here we show only the atom projected band structure of the graphene layer to highlight the Dirac cone shift over the Fermi Level. In addition, we illustrate in Fig.~\ref{fig:C-Al11O18-Al12O17-bs} (c) the contribution of the top surface states (of non-saturated atoms) to the system's electronic structure. Even though these states cross the Fermi Level, they are projected on atoms distant from the graphene layer. Hence, From the experiment perspective, these states would not strongly affect transport measurements done in the graphene layer. 

\begin{table}[!ht]
\caption{Summary of the energy and magnetic properties of the Al$_{12}$O$_{17}|$graphene interface}
\begin{tabular}{l|c|c|c}
Al$_{12}$O$_{17}|$graphene & \multicolumn{1}{l|}{Type-I''} & \multicolumn{1}{l|}{Type-II''} & \multicolumn{1}{l}{Type-III''} \\ \hline
$\Delta E_{t} $ per s.u.c. (eV)  & 4.77    & 0.00   & 3.71       \\
$\mu_{B}$ per s.u.c.             & 0.00    & 0.10   & 0.25       \\
$\mu_{B}$ of unbonded O atom     & 0.00    & 0.00   & 0.00                          
\end{tabular}\label{table:al12o17}
\end{table}

\subsubsection{The Al\texorpdfstring{$_{11}$}{TEXT} O\texorpdfstring{$_{18}|$}{TEXT}graphene interface.} Unlike the cases previously discussed, removing a single Al atom adjacent to the graphene layer restricts us from investigating only two possible geometries, namely type-I' and -II', where the type-I' is the most stable structure between them (see Table~\ref{table:al11o18}). By removing one Al atom in the geometry type-I', (three) oxygen atoms (per unit cell) at the interface contact become closer to the graphene layer. This proximity between O and C atoms forms $sp^{3}$ bonds and induces graphene to buckle. Note that its final geometry and electronic band structure is shown in Fig.~\ref{fig:C-Al11O18-Al12O17-bs}(b) is very similar to the case (type-I) where the oxide has perfect stoichiometry and oxygen is near to graphene (see Fig.~\ref{fig:C-Al12O18-NS-bs}(b)).

In the type-II' geometry, the presence of one Al in between the oxygen and graphene layer prevents all charge transfer occurs between C and O atoms. Consequently, there will be no $sp^{3}$ or structural deformation in the graphene layer. Again, this indirect interaction between O and C induces a $p$-type doping in the graphene, as shown in Fig.~\ref{fig:C-Al11O18-Al12O17-bs} (f). Notice that this electronic structure resembles that depicted in Fig.~\ref{fig:C-Al11O18-Al12O17-bs} (e), which confirms that doping may occur in a less stable system and only when oxygen atoms are near graphene but do not cause any structural damage.

\begin{table}[!ht]
\caption{Summary of the energy and magnetic properties of the Al$_{11}$O$_{18}|$graphene interface}
\begin{tabular}{l|c|c|c}
Al$_{11}$O$_{18}|$graphene & \multicolumn{1}{l|}{Type-I'} & \multicolumn{1}{l|}{Type-II'} & \multicolumn{1}{l}{Type-III'} \\ \hline
$\Delta E_{t} $ per s.u.c. (eV)  & 0.00   & 2.86    & -                        \\
$\mu_{B}$ per s.u.c.             & 1.00   & 2.79    & -                        \\
$\mu_{B}$ of unbonded O atom     & 0.47   & 0.75    & -                             
\end{tabular}\label{table:al11o18}
\end{table}

\subsection{Saturating the top surface dangling bonds}
We have previously demonstrated that breaking the bulk symmetry along the z-direction leads to a different slab surface coordination number depending on whether the interface structure is terminated with Al or O. In theory, there should be an ideal thickness or a minimum number of atomic layers for the slab that prevents the two surfaces from having a significant influence on one another. Due to experimental considerations, we restrict our analysis to a slab  $\approx 1.3~nm$ thick in this study \cite{belotcerkovtceva2022insights}. For this thickness, it is essential to investigate whether the saturation of the alumina's top surface substantially impacts the graphene band structure at the interface.

When saturating dangling bonds, we took care to ensure that the attached atom/ligand correctly fills the electron shell to maintain the same doping level; otherwise, artificial doping due to an excess or lack of electrons in the overall system may occur. To prevent this, we carefully saturated the dangling oxygen bonds of all type-II systems with hydrogen atoms, while we saturated all dangling bonds of the type-III system with hydroxide (OH-). We found that saturation of the dangling bonds restricts the number of specific examples that may be studied. For example, the type-I and type-I' geometries must not be saturated once the existing self-compensation satisfies all bonds. This brings us to five possible cases, namely type-II, -II', -II'' and type-III, -III'' geometries that will be discussed in the present section. 

\begin{figure}[t]
  \centering
  \includegraphics[width=1.0\columnwidth]{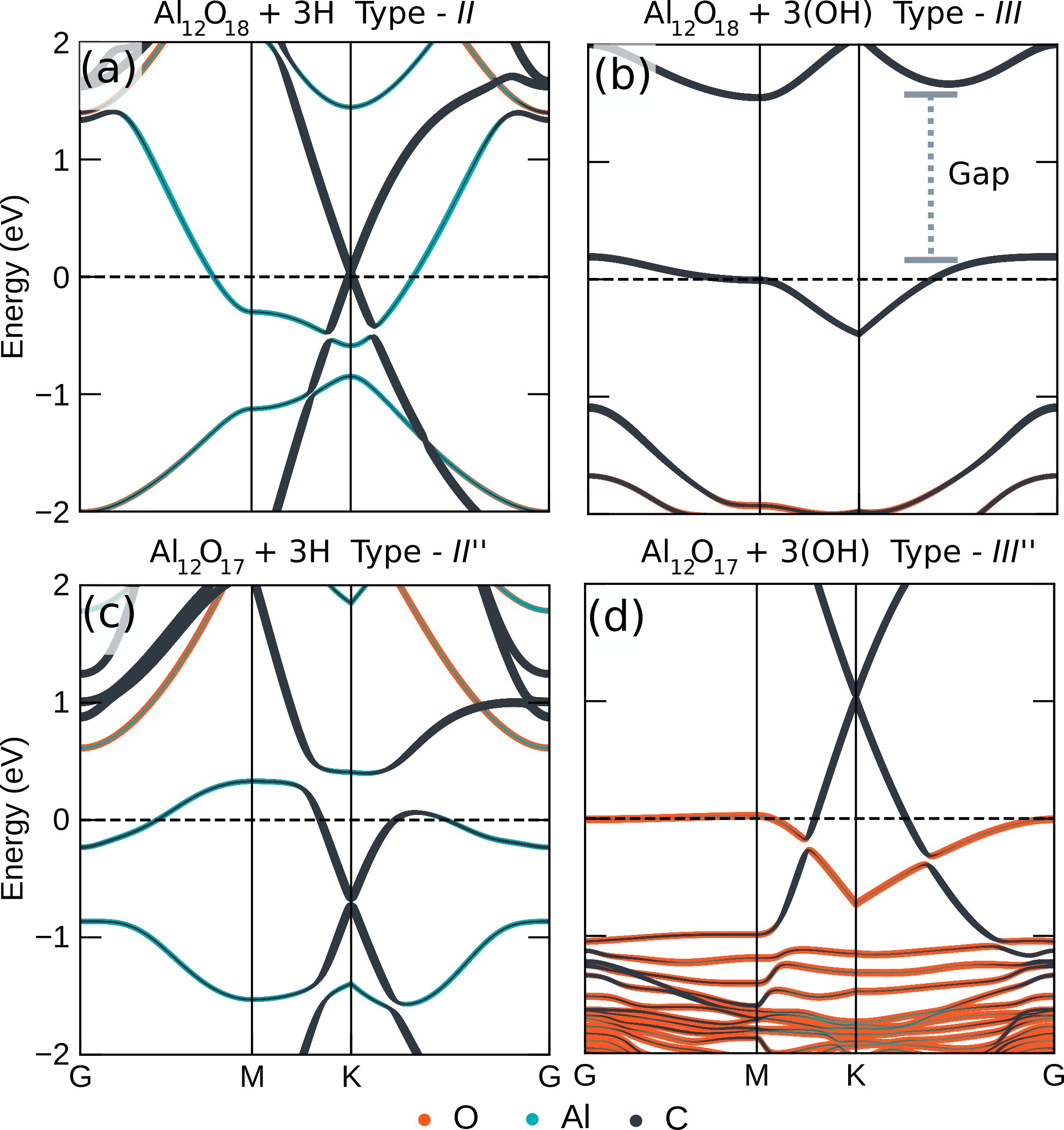}
  \caption{
  Atom projected band structure for the A$_{12}$O$_{18}$ and  A$_{12}$O$_{17}$ geometries. (a) The type-II geometry with hydrogen saturating the dangling bonds. (b) The type-III with hydroxide saturating the dangling bonds. Similarly, (c) and (d) illustrate the slightly off stoichiometric structure A$_{12}$O$_{17}$ saturated with hydrogen (type-II'') and hydroxide (type-III''), respectively.}
  \label{fig:bs_saturated_structures}
\end{figure}

Using hydrogen to saturate oxygen bonds to form hydroxide ligands the stoichiometric and off-stoichiometric of type-II, -II', -II'' and allowing full relaxation changes the final geometric structure when compared with the previous results obtained without saturating. This is partially expected because the hydrogen termination prevents Al and O atoms from the auto-compensating mechanism discussed above. However, the Al$_{12}$O$_{18}+3$H type-II and Al$_{12}$O$_{17}+3$H type-II'' also maintained their crystal structure almost intact at the contact interface near graphene, meaning that unlike the previous cases no self-compensation happens at the interface region for both structures. This has a direct influence on the electronic band structure of graphene, as shown in Fig.~\ref{fig:bs_saturated_structures}(a) and (c). Due to a lack of self-compensation at the interface contact, the charge redistribution has more substantial effects on the $\pi$-orbital of carbon atoms which causes hybridization. For the perfect stoichiometric system (Al$_{12}$O$_{18}+3$H type-II), this hybridization induces the Al conduction band states to connect with the graphene valence states of the Dirac cone. This effect becomes even more pronounced in the absence of one oxygen atom (Al$_{12}$O$_{17}+3$H type-II'') due to the electron availability increasing, and as a consequence, the central part of the Dirac cone is isolated from both valence and conduction band of the $\pi-$orbital. Due to the donating electrons coming from the Al atom, the cone is mainly pushed down below the Fermi level, suggesting that the Al atom cause $n-$type doping in the graphene monolayer when it is not fully oxidized. In the case of the Al$_{11}$O$_{18}+3$H type-II' system, however, the system shows self-compensation at the interface contact of the alumina slab, which leads to a weak interaction between the oxide and graphene, as discussed previously. Unlike the Al$_{11}$O$_{18}$ type-II' system, with no hydrogen saturation, the saturated system does not present any $sp^{3}$ bond, $p-$type doping, and the band structure is similar to the one illustrated in Fig.~\ref{fig:C-Al12O18-NS-bs}(a).

When saturating the dangling bonds of both Al$_{12}$O$_{18}$ type-III and Al$_{12}$O$_{17}$ type-III'' with hydroxide to obtain  Al$_{12}$O$_{18}+3$+(OH) type-III'' (Fig. 5b) and Al$_{12}$O$_{17}+3$(OH) type-III'' (Fig. 5d), respectively, we found that the former opens a gap of the Dirac cone, while the later keeps the $p$-type doping character first discussed in Fig.\ref{fig:C-Al11O18-Al12O17-bs}(e). For the Al$_{12}$O$_{18}+3$(OH) type-III system, the full relaxation of the atomic position shows that all oxygen atoms bind to the graphene layer, which is a different result from the previously studied case (Fig.~\ref{fig:C-Al12O18-NS-bs}(b)) where no saturation was considered and only two oxygen (out of three) per unit cell bonded to the graphene layer. For this reason, the graphene structure buckles completely, and no Dirac cone is observed in Fig.\ref{fig:bs_saturated_structures}(b). Alternatively, for the Al$_{12}$O$_{17}+3$(OH) type-III'', no $sp^{3}$ bond is formed and a $p-$type doping is observed.

\subsection{Spin polarization calculations}

As a complement to the investigation of the influence of the alumina on the graphene layer, we also explore the spin polarisation effects on the graphene electronic band structure. As discussed along with this work, the slab construction of a thin film imposes unsaturated, dangling bonds and uncompensated charges at the top surface and in the interface contact so that a spontaneous spin-polarization exists is expected. Therefore, in this section, we address the implications given by these effects.

The spin-polarized calculation can be performed by either fully relaxing the structure incorporating spin interaction or doing a self-consistent calculation that includes spins. Here, we consider the latter case. Therefore, all the discussion about the relaxed structure for the on- and the slightly-off stoichiometric system remains. We found three conditions where spontaneous magnetization emerged from self-consistent spin-polarized calculations. The first condition is when not all oxygen atoms are bonded to the graphene layer. Unlike the unbounded oxygen atom, the oxygen atoms that have filled chemical bonds are closed-shell where all electrons are paired, and no magnetization emerges. In the sections above, we have shown a few cases where only two out of three oxygen per unit cells formed a covalent bond with the graphene layer. For example, for the Al$_{12}$O$_{18}$ type-III, the calculation shows that the remaining non-bonded oxygen atom present a magnetization of \unit[0.32]{$\mu_{B}$}. The second condition relies on having an excess of oxygen in the system, which ensures that Al atoms do not saturate oxygen bonds. This case is illustrated when Al$_{11}$O$_{18}$ type-I' is on top of the graphene layer, where the unbounded oxygen atom presents a magnetization of \unit[0.47]{$\mu_{B}$}. Finally, the last condition to observe spontaneous magnetization is when we consider the type-II' structure where oxygen at the top surface can not be self-compensated. This is the case of Al$_{11}$O$_{18}$ type-II' where each top surface oxygen atom shows a magnetization of \unit[0.75]{$\mu_{B}$}. We summarize the obtained local magnetic moment per unit cell for the graphene-aluminium oxide in the Table~\ref{table:al12o18},~\ref{table:al12o17}, and \ref{table:al11o18}

\begin{figure}[t]
  \centering
  \includegraphics[width=1.0\columnwidth]{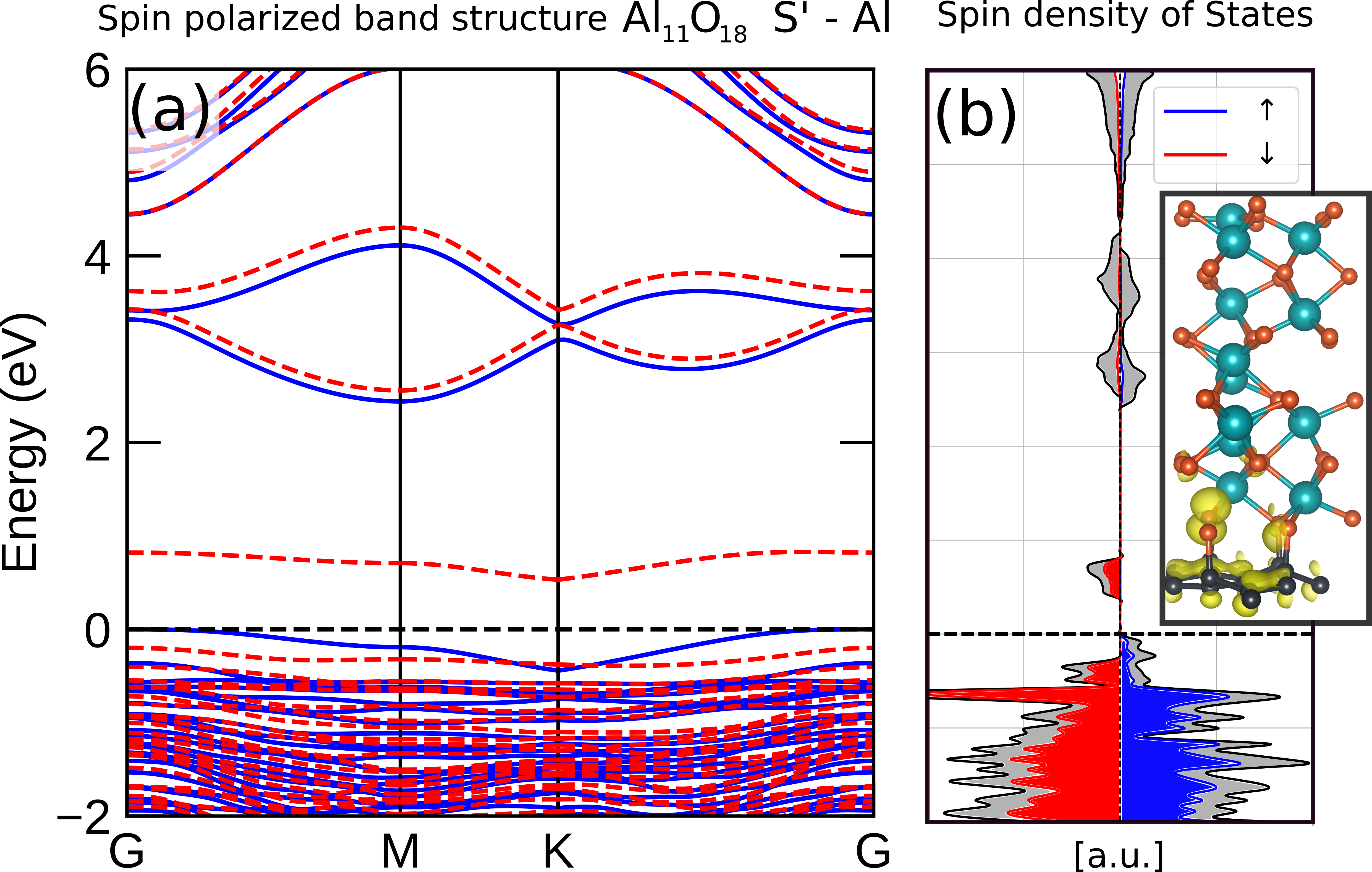}
  \caption{ (a) Spin polarized band structure and (b) spin density of states for the  A$_{11}$O$_{18}$ type-I' composite. The inset sketch shows the magnetic density plot (as a yellow sphere) around the unbonded oxygen atom and the graphene carbon atoms.}
  \label{fig:bs_and_dos}
\end{figure}

In Fig.~\ref{fig:bs_and_dos}, we show the spin density plot (inset) along with the spin projected density of state and spin-polarized band structure of the structure Al$_{11}$O$_{18}$ type-I', to illustrate the above discussion. The unbounded oxygen polarizes the carbon atoms of the graphene layer and the atoms in its close surroundings. As a result, the band structure in Fig.~\ref{fig:bs_and_dos}(a) exhibits the spin-up (blue) and spin-down (red) splitting, and  Fig.~\ref{fig:bs_and_dos}(b) emphasizes that only spin-down states are found above the Fermi level. These states are responsible for generating the spontaneous magnetization found in the unsaturated oxygen atoms and are interesting because these might explain, for example, the observed reduced spin lifetime in graphene spintronics experiments~\cite{tombros2007electronic}.

Alternatively, such spontaneous magnetization is also present in the top surface of O-terminated systems, such as all type-II geometries discussed in the previous sections. In both cases, when not saturated with hydrogen, the oxygen atoms induce a metallic behavior with a strong magnetization at the top surface. For example, the Al$_{11}$O$_{18}$ type-II' system has a magnetic moment (of about $2.79~\mu_{B}$ per unit cell) mainly due to the oxygen surface atoms. This magnitude is enough to induce a ferromagnetic state. Of course, such a particular state is unstable in our case, and a stabilization mechanism would be necessary. However, we observe that the local magnetic moment due to the oxygen atoms at the top surface would vanish if it is saturated with the charged atom (e.g., H+) or hydroxide (OH-). Therefore, this leads to an open question on how to stabilize the magnetic surface at the top of the oxide, but the answer to this question is out of the scope of this manuscript.

\section{Discussion and Conclusion}\label{sec:conclusion}

Considering practical graphene-aluminium oxide interfaces, here we simulate~\cite{belotcerkovtceva2022insights}, a thin film of aluminium oxide and study its influence on the graphene electronic band structure. We found that the graphene band structure behaves differently depending on whether Al or O is next to graphene. For some of the structures investigated here, Al and O can compensate for each other's bonds, which causes weaker interaction with graphene. For the ones in which self-compensation does not exist, we observe that a majority of oxygen atoms near graphene form $sp^{3}$ bonds when O is close to graphene. When Al is close to graphene, the oxide moves away from C atoms. We found that oxygen termination can present a significant magnetic moment for the unsaturated top surfaces. In addition, we found that depending on how the dangling bonds of the top surface are saturated, the contact interface between alumina and graphene can either interact weakly with $sp^{2}$ bonds or form an $sp^{3}$ bond between carbon and oxygens near the interface. This suggests that the mechanism of controlling the degree of $sp^{2}$ and $sp^{3}$ bond between O and C atoms at the interface is related to surface saturation. Therefore, with this study, we present insights into graphene-aluminium oxide interfaces that open up possibilities for engineering novel $sp^{3}$ controlled graphene nanoelectronic and spintronic devices.
 
The current work has practical significance in investigating the use of $AlO_{x}$ based graphene electronic and spintronic devices. Firstly, the amount of $sp^{3}$ bonds at the graphene-$AlO_{x}$ interfaces should experimentally be possible to control by the process of low-energy sputtering techniques and control of stoichiometry. Precisely controlling the density of Al or O vacancies is a challenge, but once it is optimized, it could be important for the application of graphene resistive memories and engineering locally magnetic active graphene. For instance, the $sp^{3}$ bonds can serve as carrier traps for creative resistive synaptic junctions, and controlling oxygen ions in the $AlO_{x}$ layer could enable controlling the interface covalency leading to changes in the conductivity of graphene. This could be utilized for setting and resetting processes in memristive devices\cite{huang2017graphene}. Creating magnetically active graphene could provide a feasible unconventional alternative to hydrogenated graphene, which has been seen to create exotic effects such as magnetic moments~\cite{mccreary2012magnetic} and colossal spin-orbit coupling~\cite{kim2013manipulation}, and stabilization of magnetism~\cite{giesbers2013interface}. The possibility of using chemical means, coupled with an electric field, to control the degree of metallic $sp^2$ bonds and insulating $sp^3$ bonds at the alumina graphene interface could also open up for a large on-off ratio of graphene transistor applications. 

\section{Acknowledgment}

We are grateful to I.P. Miranda for the very fruitful discussions that were stimulating to write this paper. 
This work was supported by Swedish Research Council (VR), Swedish National Infrastructure for computing (SNIC), European Research Council (ERC), the Kunt and Alice Wallenberg foundation (KAW), The Swedish Energy Agency (Energimyndigheten), StandUPP and eSSENCE.  

\bibliography{ref}
\end{document}